\documentclass[sigconf]{acmart}

\AtBeginDocument{%
  }

\copyrightyear{2025}
\acmYear{2025}
\setcopyright{acmlicensed}\acmConference[CHI '25]{CHI Conference on Human Factors in Computing Systems}{April 26-May 1, 2025}{Yokohama, Japan}
\acmBooktitle{CHI Conference on Human Factors in Computing Systems (CHI '25), April 26-May 1, 2025, Yokohama, Japan}
\acmDOI{10.1145/3706598.3714198}
\acmISBN{979-8-4007-1394-1/25/04}

\usepackage{enumitem}
\usepackage{float}
\usepackage{xcolor} 

\begin{document}



\title[Human Creativity in the Age of LLMs]{Human Creativity in the Age of LLMs}
\subtitle{Randomized Experiments on Divergent and Convergent Thinking}

\author{Harsh Kumar}
\email{harsh@cs.toronto.edu}
\orcid{0000-0003-2878-3986}
\affiliation{%
  \institution{University of Toronto}
  \city{Toronto}
  \state{Ontario}
  \country{Canada}
}

\author{Jonathan Vincentius}
\email{jon.vincentius@mail.utoronto.ca}
\orcid{0009-0007-2443-1576}
\affiliation{%
  \institution{University of Toronto}
  \city{Toronto}
  \state{Ontario}
  \country{Canada}
}

\author{Ewan Jordan}
\email{ewan.jordan@mail.utoronto.ca}
\orcid{0009-0001-3176-5036}
\affiliation{%
  \institution{University of Toronto}
  \city{Toronto}
  \state{Ontario}
  \country{Canada}
}

\author{Ashton Anderson}
\email{ashton@cs.toronto.edu}
\orcid{0000-0003-3089-6883}
\affiliation{%
  \institution{University of Toronto}
  \city{Toronto}
  \state{Ontario}
  \country{Canada}
}

\renewcommand{\shortauthors}{Harsh Kumar et al.}

\begin{abstract}

Large language models are transforming the creative process by offering unprecedented capabilities to algorithmically generate ideas. While these tools can enhance human creativity when people co-create with them, it's unclear how this will impact unassisted human creativity. We conducted two large pre-registered parallel experiments involving 1,100 participants attempting tasks targeting the two core components of creativity, divergent and convergent thinking. We compare the effects of two forms of large language model (LLM) assistance---a standard LLM providing direct answers and a coach-like LLM offering guidance---with a control group receiving no AI assistance, and focus particularly on how all groups perform in a final, unassisted stage. Our findings reveal that while LLM assistance can provide short-term boosts in creativity during assisted tasks, it may inadvertently hinder independent creative performance when users work without assistance, raising concerns about the long-term impact on human creativity and cognition.
\end{abstract}


\begin{teaserfigure}
  \includegraphics[width=\textwidth]{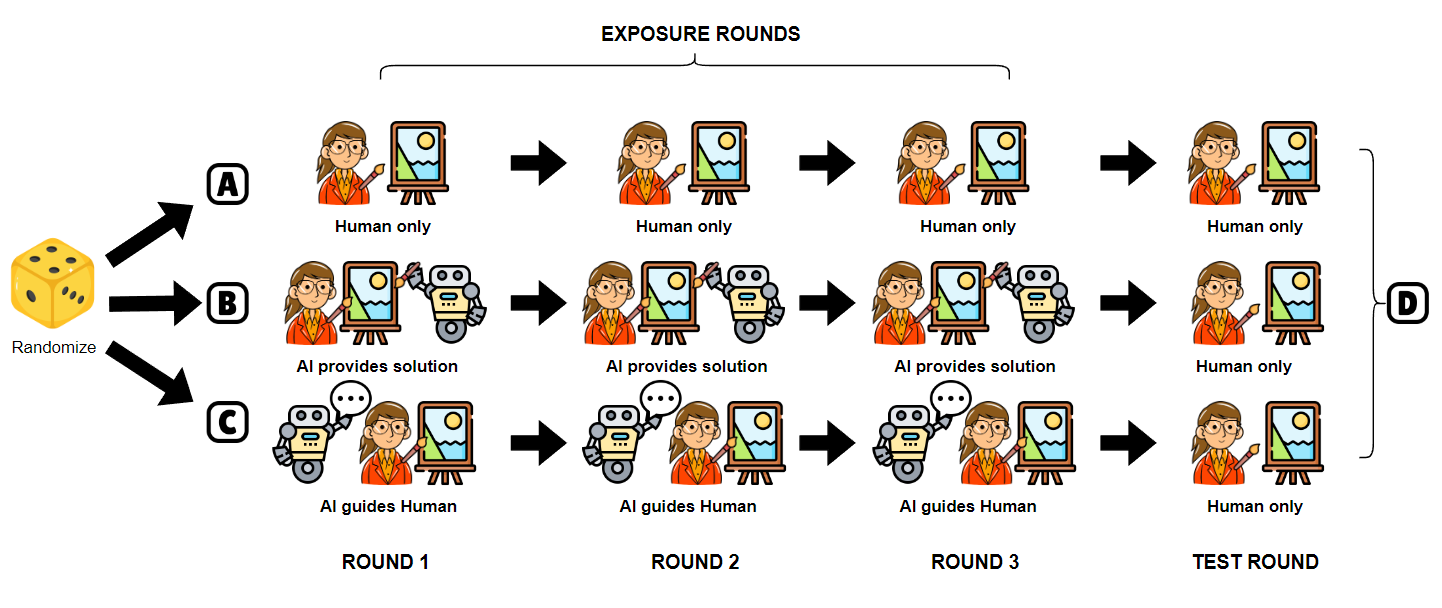}
  \caption{Experimental framework for measuring the impact of AI use on Human creativity. Participants engage in a series of \textit{Exposure} rounds where they are randomized to either receive - \textbf{(A)} No assistance, \textbf{(B)} LLM solution (standard): This could be analogous to using a chat LLM such as ChatGPT for the task, or \textbf{(C)} LLM guidance (coach-like): In this case, participants receive response from a customized LLM which guides them through the creative process. Finally, in the last round, all participants are asked to do the same creative task without any assistance as a \textit{Test}. \textbf{(D)} The performance and creative outputs in this unassisted round are the primary measures for evaluating the impact of using LLMs on Human cognition.}
  \Description{Figure 1 depicts a pictorial flow chart illustrating the high level structural overview of the experiments performed. Participants are randomized into one of three conditions for three exposure rounds: A (human-only), B (AI provides full solutions), and C (AI guides but doesn't provide full solutions). In Condition A, participants complete 3 tasks independently. In Condition B, participants complete 3 tasks with AI giving the solution, while in Condition C, participants complete 3 tasks with AI guidance without providing the final answer. In the final test round (D), all participants work independently, regardless of their previous condition.}
  \label{fig:teaser}
\end{teaserfigure}


\maketitle

\section{Introduction}
\label{sec:intro}
The rise of generative artificial intelligence tools such as ChatGPT has the potential to upend the creative process. By their nature as \emph{generative} systems, they offer an unprecedented capacity to algorithmically generate ideas, and perhaps even to create. State-of-the-art generative AI systems have reached proficiency levels that not only match human creativity in certain evaluations \cite{koivisto2023best, epstein2023art}, but can also enhance the creative output of knowledge workers \cite{anderson2024homogenization, agres2015conceptualizing, hsueh2024counts}. We are living through a time of creative transformation, with AI visual art, AI music, and AI-enhanced videos rapidly proliferating. 

But what does the era of generative AI hold for human creativity? When AI assistance becomes a regular part of creative processes, will human creativity change? There are natural concerns that reliance on generative AI might impair an individual's inherent ability to think creatively without assistance. Further, there is preliminary evidence that widespread dependence on similar generative AI tools could lead to a homogenization of thought. This could in turn reduce the diversity that drives collective innovation and stifle breakthroughs across various fields. However, there is also reason to believe that AI assistance could spark human creativity. Working with a fresh creative partner might be similarly stimulating regardless of whether the partner is human or machine. If this is the case, AI could prove to be a force for the flourishing of human creativity. Understanding how and whether co-creating with AI affects an individual's ability to generate creative ideas independently is therefore of critical importance. 


This is a complex challenge. Creativity is inherently subjective and difficult to measure, making it hard to assess changes in an individual's creative ability. Factors like prior knowledge, experience, and environmental context all interplay with creativity, making it hard to isolate the effects of AI usage ``in the wild''. The rapid evolution of AI technologies further complicates matters, as their capabilities and effects are constantly changing. Without clear methods to evaluate these impacts, we cannot fully grasp how AI integration might alter the creative landscape.


In this work, we investigate the impact of Large Language Models (LLMs) assistance on human creativity by examining the two fundamental components of creativity: divergent and convergent thinking. Divergent thinking involves the generation of multiple, unique ideas, fostering exploration and innovation \cite{runco1991divergent, runco2012divergent}. Convergent thinking, in contrast, focuses on refining these ideas to select the most effective solutions \cite{baer2014creativity, guilford1967creativity}. We investigate how different forms of LLM assistance influence these cognitive processes by comparing two types of LLMs—a standard LLM that provides direct answers out of the box, and a coach-like LLM that offers guidance and prompts to stimulate thinking—in contrast to a control condition with no LLM assistance.

Specifically, we address the following research questions:
\vspace*{0.5em}
\begin{description}
    \hrule height 0.1em
    \item[RQ1] How do standard LLM assistance and coach-like LLM guidance, compared to no assistance, affect an individual's divergent thinking abilities when generating creative ideas independently?
    \item[RQ2] What are the impacts of standard LLM assistance and coach-like LLM guidance, versus no assistance, on an individual's convergent thinking skills in independently refining and selecting ideas?
\end{description}
\hrule height 0.025em
\vspace*{0.5em}

To answer these questions, we designed and conducted two pre-registered parallel experiments with 1,100 participants that assess how these forms of LLM assistance influence unassisted human creativity, compared to a control group with no AI assistance. Figure \ref{fig:teaser} illustrates the high-level design of the experiments. In both experiments, participants were randomly assigned to one of three conditions: standard LLM assistance, coach-like LLM guidance, or no assistance (control). They engaged in a series of \emph{exposure} rounds in which they completed creative tasks using their assigned form of LLM assistance. After a delay period, participants then completed the same type of creative tasks unassisted in the \emph{test} rounds. This design allowed us to examine both the immediate effects of LLM assistance during the exposure rounds and the residual effects on unassisted creative performance during the test rounds.

For divergent thinking, we utilized the Alternate Uses Test (AUT), where participants were asked to come up with creative uses for common objects. We found that exposure to LLM assistance—whether providing ideas or strategies—did not enhance participants' originality or fluency in subsequent unassisted tasks. In some cases, it even led to decreased originality and reduced diversity of ideas, suggesting a potential homogenization effect where individuals generate more similar ideas after using LLM assistance. We employed the Remote Associates Test (RAT) for convergent thinking, which requires finding a word that connects three given words. Our findings indicate that while LLM assistance improved performance during the assisted tasks, it did not translate into better performance in subsequent unassisted tasks. Participants who received guidance from LLMs performed worse in the unassisted test rounds compared to those with no prior LLM exposure.

The paper contributes empirical findings on the impact of LLM assistance on human creativity, specifically focusing on divergent and convergent thinking. Specifically, our study (1) provides empirical evidence that LLM assistance boosts performance during assisted tasks but may hinder independent creative performance in unassisted tasks; (2) reveals the differential impacts of LLMs on divergent and convergent thinking, highlighting users' skepticism toward LLM assistance in divergent tasks and beneficial effects in convergent tasks; and (3) identifies persistent homogenization effects due to LLM-generated strategies, posing challenges for designing effective LLM coaching systems. The paper also offers design implications for developing LLM-based tools that enhance human creativity without undermining independent creative abilities.

\section{Related Work}
\label{sec:background}

This paper builds on a rich body of literature on creativity, LLMs, and the impact of generative AI on human cognition. Although previous studies have laid the groundwork in these fields, our work makes novel contributions toward understanding the impact of different forms of LLM assistance on human creativity.

\subsection{Theories of Human Creativity}

Divergent and convergent thinking constitute critical components of the creative process \cite{goldschmidt2016linkographic, zhang2020metacontrol, cortes2019re, de2019scientific, jaarsveld2012creative, razumnikova2020divergent}. \textbf{Divergent thinking} involves generating a wide range of ideas, exploring multiple possibilities, and embracing unconventional approaches \cite{runco1991divergent, runco2012divergent, baer2014creativity, guilford1967creativity}. In contrast, \textbf{convergent thinking} focuses on narrowing down these ideas, selecting the most viable options, and refining them into coherent solutions. Using LLMs can differentially influence these processes, with distinct immediate and long-term effects on divergent and convergent thinking. We draw on key theories by Poincaré and Boden to frame our investigation \cite{gaut2003creativity, livingston2009chapter, boden2010creativity}. Poincaré’s four-phase model encapsulates creativity as both an unconscious and an active process. Divergent thinking is vital during the \textit{Preparation} and \textit{Incubation} phases, where the mind explores various possibilities \cite{baer2014creativity, gilhooly2013incubation}. Convergent thinking becomes crucial in the \textit{Insight} and \textit{Revision} phases, refining ideas into viable solutions \cite{cropley2006praise, simonton2015praising}. This model underscores that creativity involves not only generating numerous ideas, but also selecting and refining them \cite{zhu2019convergent, hommel2011bilingualism}.

Boden’s theory further introduces measures of creativity, distinguishing between \textbf{P-creativity} (psychological creativity) and \textbf{H-creativity} (historical creativity) \cite{boden2007creativity, boden2010creativity}. P-creativity refers to ideas novel to the individual, while H-creativity refers to ideas novel within the broader context of human knowledge. This distinction allows us to assess creativity on both a personal and historical scale. Divergent thinking can be hypothesized to contribute primarily to P-creativity, where the generation of new ideas is the key \cite{runco1991divergent, runco2010divergent, runco2012divergent}. In contrast, convergent thinking is essential for advancing these ideas toward H-creativity, where their novelty and value must be recognized within a broader context.

Our experiments explore how LLMs influence these aspects of creativity. To investigate divergent thinking, we utilize the Alternate Uses Test (AUT), where participants are prompted to generate as many creative uses as possible for a common object \cite{guilford1967creativity, gilhooly2007divergent}. For convergent thinking, we use the Remote Associates Test (RAT), where participants find a single word that connects three seemingly unrelated words, assessing their ability to converge on a correct and meaningful solution \cite{mednick1968remote, wu2020systematic, lee2014measure}.

\subsection{LLMs as Tools for Creativity}
LLMs have demonstrated notable creative performance, often performing as well as, or even better than, average humans in various creative tasks. However, they still fall short when compared to the best human performers, particularly experts \cite{koivisto2023best, hubert2024current}. For instance, Chakrabarty \textit{et al.} \cite{chakrabarty2024art} found that while LLMs performed competently in creative writing tasks, professionals consistently outperformed them. Anderson \textit{et al.} \cite{anderson2024homogenization} observed that ChatGPT users generated more ideas than those in a control condition. However, these ideas tended to be homogenized across different users, suggesting a limitation in the diversity of creative output when LLMs are used. Koivisto \textit{et al.} \cite{koivisto2023best} compared LLM vs. human performance in a divergent thinking task, but without examining how LLM use affects human creativity. Ashkinaze \textit{et al.} \cite{ashkinaze2024ai} conducted a single round idea generation with LLM support and asked participants to provide a single original idea. These studies are particularly relevant when LLMs are employed to perform entire tasks on behalf of users.

Beyond performing tasks autonomously, LLMs can also be utilized to guide users through the \textit{conceptual spaces} of creative thinking, serving as tools for `creativity support' \cite{agres2015conceptualizing, hsueh2024counts, qin2024character}. This approach positions LLMs not merely as replacements for human creativity but as enhancers of the creative process, helping users navigate and explore creative possibilities more effectively. The literature on Human-AI collaboration in creative tasks is rapidly growing. Lee \textit{et al.} \cite{lee2022coauthor} introduced a dataset for analyzing GPT-3's use in creative and argumentative writing, suggesting that the HCI community could foster more detailed examinations of LLMs' generative capabilities through such datasets. Suh \textit{et al.} \cite{suh2024luminate, suh2023sensecape} highlighted that the current interaction paradigm with LLMs tends to converge on a limited set of ideas, potentially stifling creative exploration. They proposed frameworks that facilitate the exploration of structured design spaces, allowing users to generate, evaluate, and synthesize many responses.

A key question remains: what happens to human creativity, particularly cognition, when humans repeatedly use LLMs for creative tasks—either directly to generate ideas or solutions, or as guides through the creative process? While there are concerns that this could lead to a deterioration of creative abilities, there is also optimism that these tools, if properly designed, could enhance human creativity while providing momentary assistance \cite{epstein2023art, heersmink2024use}. However, empirical evidence to answer this question definitively is still lacking.

\begin{figure*}
\centering
\includegraphics[width=\textwidth]{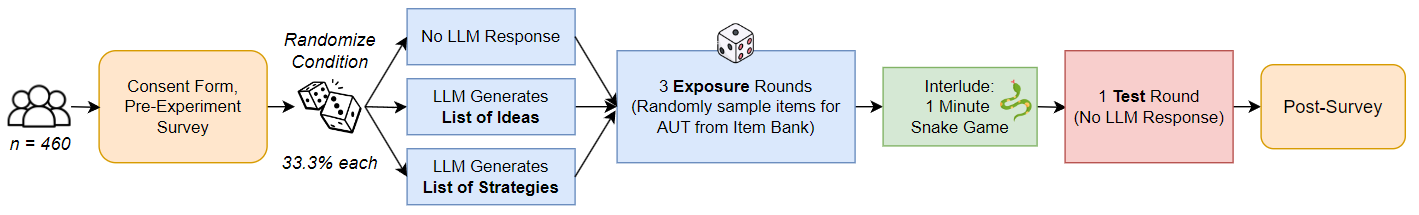}
\caption{Schematic of design for Experiment 1 on divergent thinking.}
\Description{Figure 2 illustrates the high level structural overview of the divergent thinking experiment. 460 participants start with a consent form and pre-experiment survey before being randomized into one of three conditions for three exposure rounds. These conditions are: A (human-only), B (AI provides a list of ideas), and C (AI provides a list of brainstorming strategies). The items provided in these exposure rounds are randomly sampled from an item bank. After the exposure rounds, participants engage in a 1 minute snake game as a brief interlude before moving onto a test round that all participants carry out without any AI support. Finally, they end with a post-survey.}
\label{fig:exp-design-1}
\end{figure*}

\subsection{Generative AI and Human Cognition}
The continued use of Generative AI is significantly impacting our society, particularly in how \textit{culture} is created and propagated \cite{brinkmann2023machine, ashkinaze2024ai}. These technologies are reshaping the production of cultural artifacts and the means through which culture is disseminated and experienced. The influence of Generative AI extends to our cognitive abilities, with effects that may be polarizing \cite{heersmink2024use}. On the one hand, the pervasive use of AI tools might lead to a massive homogenization of creative output, a trend that could persist even after humans stop using these tools \cite{anderson2024homogenization}. This raises concerns about the potential stifling of creativity and the reduction of diversity in thought and expression. On the other hand, Generative AI can unlock unprecedented creative growth and learning, allowing users to expand their cognitive horizons and engage in innovative forms of thought. 

Hofman \textit{et al.} \cite{hofman2023steroids} introduced a sports metaphor to conceptualize the spectrum of the impact of Generative AI on human cognition. They describe three distinct roles that AI can play: steroids, sneakers, and coach. ``Steroids'' represent AI as a tool that provides short-term performance gains, but with potential long-term detrimental effects. ``Sneakers'' symbolize AI tools that augment human skills without long-term adverse consequences. Lastly, the ``Coach'' role reflects AI as a guide that helps individuals improve their own capabilities, extending beyond immediate assistance to foster long-term cognitive growth. Collins \textit{et al.} \cite{collins2024building} proposed AI systems as `thought partners,' designed to meet human expectations and complement our cognitive limitations. They outline several modes of collaborative thought in which humans and AI can engage, drawing insights from computational cognitive science to suggest how these partnerships can enhance human thinking.

\paragraph{Situating work in broader HCI literature.} Although the long-term effects of Generative AI use has been studied empirically in domains such as education \cite{kumar2023math, bastani2024generative}, web search \cite{spatharioti2023comparing}, etc., there is a lack of empirical evidence on the effect of Generative AI tools on our convergent and divergent thinking abilities. 
Our approach differs significantly from previous work that focuses solely on measuring and improving the creative output of LLMs themselves \cite{lu2024llm}, without considering how their use affects human (user) creativity. Unlike previous work, we examine and discuss both aspects of creativity (divergent and convergent thinking) together for a comprehensive understanding of how LLMs shape human creative processes. Prior studies have used AUT and RAT to evaluate the creative abilities of LLMs \cite{lu2024llm, koivisto2023best} or to assess Human+AI creativity under a single exposure \cite{ashkinaze2024ai}. In contrast, our work approximates real-world interactions by incorporating multiple sequential exposures to LLMs. This allowed us to capture how users adapt their creative processes over time, providing a more realistic assessment of the impacts of LLM-assisted creativity. In addition, we experimentally measure the impact of these repeated exposures on \textit{human} creativity (residual effects of using LLMs), an aspect missing from existing research on LLM and creativity.

\section{Experiment 1: Divergent Thinking}
A critical aspect of creativity is the ability to generate a wide range of high-quality ideas, often called divergent thinking \cite{runco1991divergent, guilford1967creativity}. LLMs have the capacity to generate a significant quantity of ideas, often exceeding what an individual can produce, without the constraints of time or context. However, the impact of using LLM-generated ideas on an individual’s ability to think divergently and come up with ideas independently remains underexplored. In addition to producing ideas, LLMs can offer structured frameworks that guide users in their creative processes, much like a coach \cite{baer2014creativity}. To understand these dynamics, our first pre-registered\footnote{\url{https://aspredicted.org/dkvp-n785.pdf}} experiment investigates the effects of LLMs that either directly provide ideas or guide participants using a framework, helping them to develop their own ideas.

\subsection{Experimental Design} 
We designed a three-condition, between-subjects experiment to understand how participants’ unassisted divergent thinking is affected by the presence of LLM assistance during prior divergent thinking tasks (see Figure \ref{fig:exp-design-1}). The study was approved by the ethics board of the local university. We employed the Alternate Uses Test (AUT), which is the most widely used divergent thinking task \cite{guilford1967creativity}. Participants in this task were asked to come up with novel and creative uses for common everyday objects, outside of their intended use. They were told ``The goal is to come up with creative ideas, which are ideas that strike people as clever, unusual, interesting, uncommon, humorous, innovative, or different. Your ideas don’t have to be practical or realistic; they can be silly or strange, even, so long as they are CREATIVE uses rather than ordinary uses…’’. For instance, an alternate use of pants might be as a makeshift flag.

\begin{figure*}
\centering
\includegraphics[width=\textwidth]{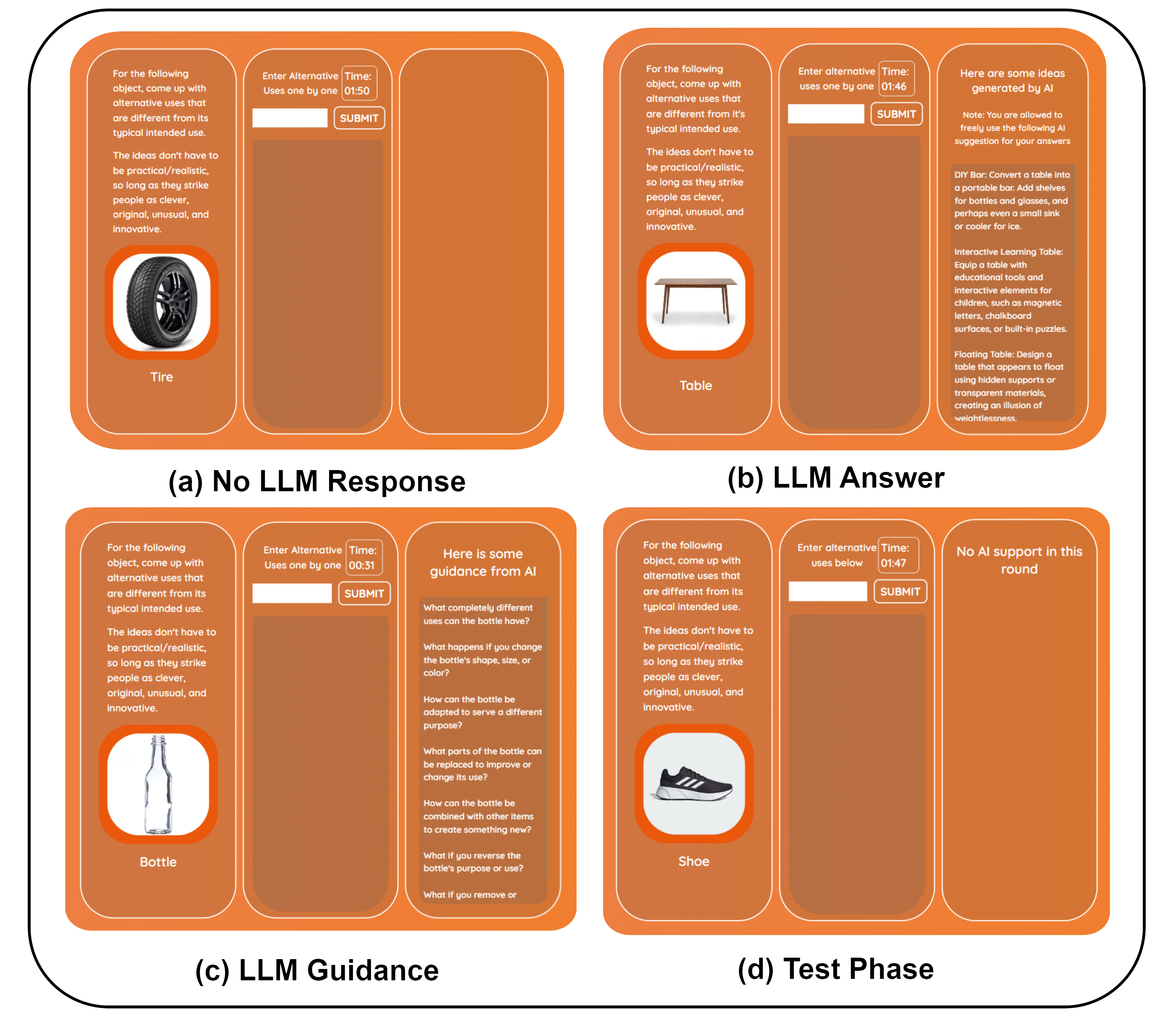}
\caption{Interface used in the divergent thinking experiment across all 3 exposure round conditions and test rounds.}
\Description{Figure 3 shows the 4 user interfaces used in the divergent thinking experiment, (No LLM Response, LLM Answer, and LLM Guidance exposure conditions, as well as for the test round). All screens are split into 3 columns: the left column contains item from the given AUT problem as well as instructions, the middle column contains a text input box at the top for entering alternate uses, then a list of all uses written so far, and the right column contains the respective AI assistance.}
\label{fig:exp-screens-1}
\end{figure*}

\subsubsection{Experimental conditions and treatment}
Our experimental design involved two main phases: an exposure (paired) phase during which participants attempted the task with an LLM partner, and a test (solo) phase during which participants attempted the task on their own. Participants were given two-minute time frames per object to submit their ideas one by one, and could freely edit or delete any previously submitted ideas. We control for time per task to emulate real-world scenarios where knowledge workers will have the same deadline for a task, irrespective of whether they are using LLM or not. During the exposure phase, participants are introduced to three objects, one after the other. Every participant was randomly assigned to receive one of three types of LLM responses:

\begin{tabular}{@{}p{1.5cm}p{6.2cm}@{}}
    \textit{None:} & Control group. No LLM support provided. \\
    \textit{List of Ideas:} & GPT-4o generated a list of alternate uses for the given object. Seven randomly sampled uses were shown to the participant, which they could freely use in their responses. In other words, an LLM attempted the task and shared its responses with the participant. \\
    \textit{List of Strategies:} & GPT-4o, using a specialized system prompt (shown in Figure \ref{fig:pre-prompt1}), generated seven strategies based on the SCAMPER technique \cite{serrat2017scamper}. In this condition, an LLM provided guidance to the participant but refrained from sharing explicit answers to the task. \\
\end{tabular}

The \textit{List of Strategies} condition was designed to capture a common real-world use of LLMs: obtaining structured thinking or brainstorming frameworks rather than direct answers. Although this approach may introduce additional cognitive load during the \textit{exposure} phase, it can potentially foster deeper, more enduring benefits observed during the \textit{test} phase—an effect analogous to the long-term learning advantages reported in educational studies leveraging LLMs \cite{kumar2023math, bastani2024generative}. The assigned type of LLM response appeared 5 seconds after the item was shown to the participant, and appeared character by character, similar to other chat LLM interfaces. Figure \ref{fig:exp-design-1} shows the schema of the experiment design. Following the exposure phase, participants engaged in a brief distractor task to simulate forgetting, playing a game of Snake for one minute. In this subsequent phase, participants were assigned the Alternate Uses Task for a new object selected at random, this time without LLM support. None of the participants in any condition had LLM support in the \textit{test} round. This was done to measure the residual effects of the use of LLMs in previous \textit{exposure} rounds.


Participants completed a pre-survey before beginning the experiment and a post-survey after completing the test phase, where we collected self-perceived creativity levels and their attitudes toward AI, along with other subjective measures (such as perceived difficulty of the test round, any strategies they utilized, and if there were any technical issues).


\subsubsection{Stimuli}
The items in each round were randomly sampled from five items: tire, pants, shoe, table, bottle. We chose these five particular objects as the originality scoring measure had the highest correlation with human judgements for these five objects \cite{organisciak2023beyond}. Figure \ref{fig:exp-screens-1} shows the different responses shown to participants in different conditions. 

\subsubsection{Dependent Variables}
AUT allows us to measure different dimensions of divergent thinking. As such, using LLMs may impact each of these dimensions differently. These dimensions include:

\begin{itemize}
    \item \textbf{Originality (How original the idea is):} We measure the originality of each AUT idea with an existing fine-tuned GPT-3 classifier \cite{organisciak2023beyond}. The model was fine-tuned with human judgments of AUT originality (where human raters judged, on a scale of 1 to 5, the originality of an idea given the object) and achieved an $r=0.81$ overall correlation with human judgments. For our experiment, we chose the five items that had the highest accuracy for the model ($r>0.88$).
    
    \item \textbf{Fluency (How many ideas):} We measure fluency by counting the number of ideas the participant generated in a given round.
    
    \item \textbf{Individual-level diversity of idea set (How different are ideas compared to each other for a given object):} We first embed all ideas using SBERT \cite{reimers2019sentence}. Idea diversity is the median pairwise cosine distance between idea embeddings in the idea set. This is a complementary measure to Originality. While Originality is a property of the idea, Diversity is a property of the idea set, so it is possible to have a diverse set of non-creative ideas if each individual idea is not creative (by originality metric) but different from one another \cite{ashkinaze2024ai}.
    
    \item \textbf{Creative flexibility (How different are ideas in the Test round compared to the Exposure rounds):} Creative flexibility allows participants to switch between different concepts and perspectives. This is also sometimes referred to as \textbf{p-creativity}. In the framework of our experiment, we operationalize flexibility by checking how similar the ideas generated in the Test round are to the ideas generated in the Exposure rounds. We measure this by finding the maximum cosine similarity between the embedding of an idea in the Test set and all the ideas in the Exposure set. We use the maximum rather than a measure of central tendency because if a participant is inspired by an idea, it would likely be a single idea \cite{roemmele2021inspiration}. 

    \item \textbf{Group-level diversity of ideas (How different are the ideas generated by participants across the group for a given phase):} The goal of this analysis was to simulate how individuals contribute to idea generation as a group, with a focus on understanding the potential homogenization of ideas at the group level. Specifically, we aimed to study whether LLM assistance leads to a convergence of ideas, reducing diversity within a group. To achieve this, we performed 150 Monte Carlo runs, where we sampled 7 ideas per round, simulating group contribution dynamics. We calculated the median pairwise cosine distances between the embeddings of these ideas, using SBERT to encode them. This sampling process reflects the number of participants in each condition. The median pairwise cosine distance across each Monte Carlo run was used to evaluate diversity for each phase and condition. This approach allowed us to assess whether different forms of LLM assistance (e.g., providing direct ideas or acting as a coach) promote homogenization or help maintain diversity of ideas at the group level, even after stopping to use LLMs.
\end{itemize}

\begin{figure*}
\centering
\includegraphics[width=0.75\textwidth]{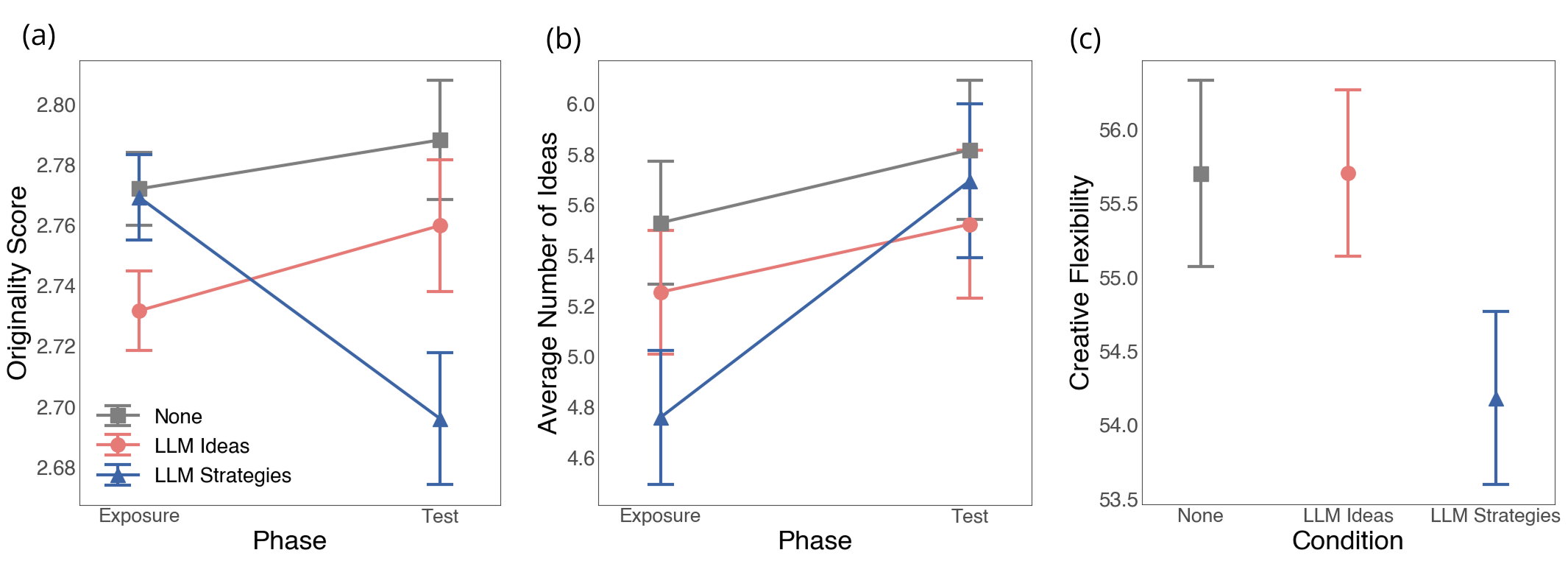}
\caption{Plots of the Alternate Uses Task ideas for the various divergent thinking dimensions (Segmented by phase and/or LLM response type). The left figure shows idea originality scores, the middle figure indicates idea fluency, and the right figure presents participant creative flexibility.}
\Description{Figure 4 contains plots of the Alternate Uses Task ideas for the various divergent thinking dimensions (Segmented by phase and/or LLM response type). The left figure shows originality score from 2.68 to 2.80 on the Y-axis, and phase(Exposure and Test) on the X-axis. No LLM response shows an increase from 2.77 to 2.78 from Exposure to Test, LLM Answer shows an increase from 2.73 to 2.75, and LLM Guidance shows a steep decrease from 2.77 to 2.69. The middle figure shows the Average Number of Ideas from 4.6 to 6.0 on the Y axis and the phase(Exposure and Test) on the X axis. No LLM Response shows an increase from 5.5 to 5.7 across the phases, LLM Answer shows a slight increase from 5.2 to 5.4, and LLM Guidance has a steep increase from 4.7 to 5.6. Error bars for all data points are roughly 0.5 wide. The right figure shows Creative Flexibility from 53.5 to 56.0 on the Y axis and Condition (No LLM Response, LLM Answer, LLM Guidance) on the X Axis. No LLM Response and LLM Answer are 55.7 and LLM Guidance is 54.1, with error bars being roughly 1 unit long.}
\label{fig:divergent-results-1}
\end{figure*}

\begin{figure*}
\centering
\includegraphics[width=0.5\textwidth]{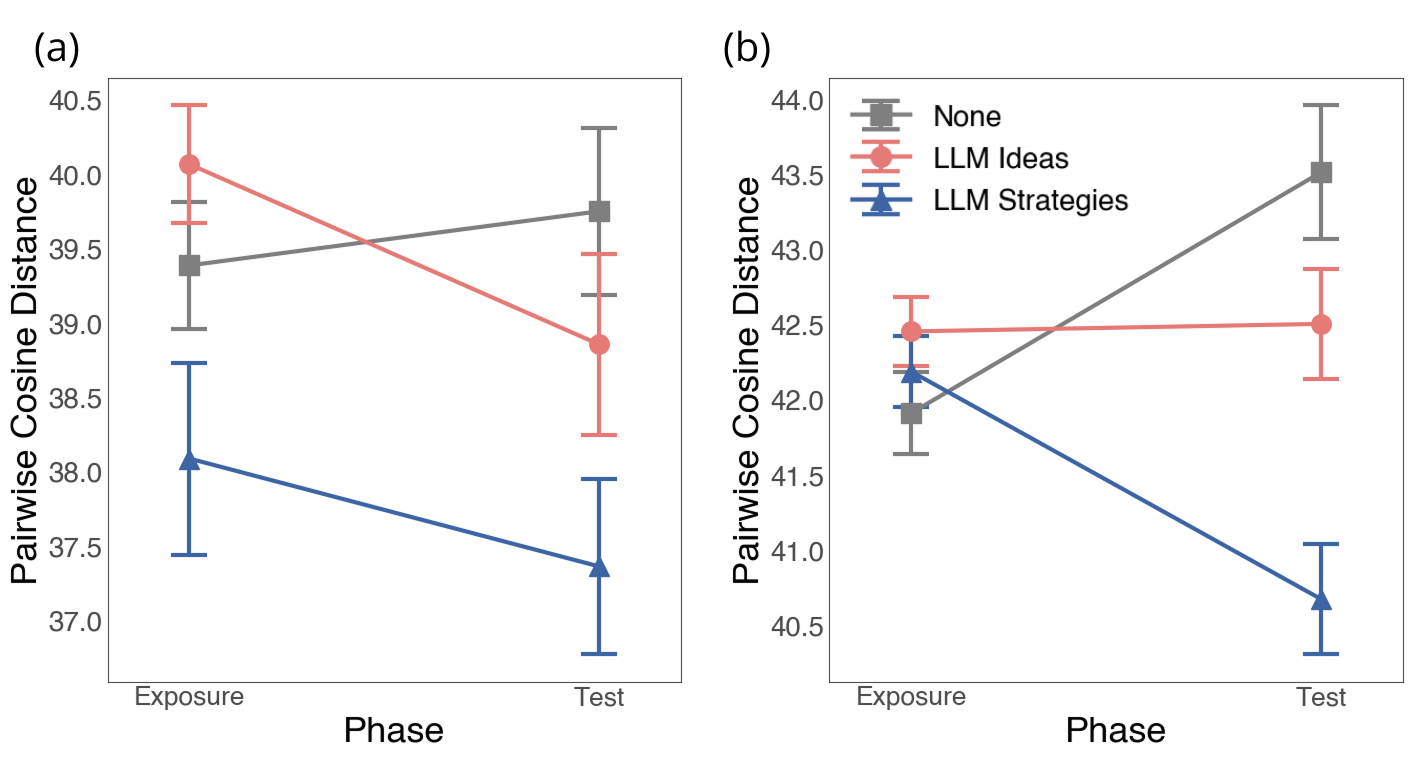}
\caption{Plots of the average individual- (left) and group-level (right) median diversity, segmented by experiment conditions and phases. Higher values denote more difference between ideas. Error bars represent $\pm$ one standard error of mean.}
\Description{Figure 5 shows two side-by-side plots of the average individual- and group-level median diversity, segmented by experiment conditions and phases. Higher values denote more difference between ideas. The average individual-level median diversity plot has Pairwise Cosine Distance from 37 to 40.5 on the Y axis. No LLM Response shows a slight increase from 39.4 to 39.7, LLM Answer shows a decrease from 40.0 to 38.8, and LLM Guidance shows a decrease from 38.0 to 37.3 (all from exposure phase to test phase). The average group-level median diversity plot has Pairwise Cosine Distance from 41 to 44 on the Y axis. No LLM Response shows an increase from 41.5 to 42.8, LLM Answer shows an increase from 42.4 to 43.4, and LLM Guidance shows a decrease from 41.7 to 41.2 (all from exposure phase to test phase).}
\label{fig:divergent-results-2}
\end{figure*}

\subsubsection{Analysis}
We follow a long tradition of scoring responses to the AUT computationally \cite{beaty2021automating, beaty2022semantic}. Following our pre-registration, we conducted a Kruskal-Wallis H test to compare the distribution of originality across the three conditions: None, List of Ideas, and List of Strategies, using a significance level of 0.05. In the event of a significant Kruskal-Wallis result, Dunn’s test with Bonferroni correction was applied as a post-hoc test for pairwise comparisons between the conditions.

This same analysis procedure was applied for other dependent variables, including individual-level and group-level diversity, fluency, and creative flexibility. We report the significance of the overall Kruskal-Wallis H test and the results of Dunn’s test for specific pairwise comparisons for each measure.

\subsubsection{Participants}
We recruited 460 participants from Prolific. Based on a power analysis using simulated and pilot data, we determined that this sample size will be necessary to achieve 70\% power with a moderate effect size and a significance level of 0.05. The overall experiment took around 12 minutes to complete and participants were paid \$1.57. Participants were based in the US or UK, and fluent in English. On average, participants felt they were more creative than 48.12\% of the population, at the start of the experiment.

\subsection{Results}
We report on the analysis of 9,457 ideas generated by participants across all conditions and phases. 

\subsubsection{Originality} 
Figure \ref{fig:divergent-results-1}a shows the average originality of ideas across conditions. In the \textit{Exposure} phase, the mean originality was similar across conditions, as indicated by the Kruskal-Wallis H test (\textit{H}(2) = 3.78, \textit{p} = 0.151). Interestingly, participants who received the \textit{List of Ideas} performed slightly worse than other conditions, despite being exposed to LLM ideas with a mean originality of 3.25 ($\pm0.1$). This is nearly 0.5 points higher than the average originality of ideas generated by participants in this condition (on a scale of 1 to 5). This discrepancy suggests that participants may struggle to accurately assess the quality of AI-generated ideas, or perhaps, when provided with AI ideas, they may prioritize coming up with their own, potentially lower-quality ideas. These findings highlight the need for designing effective reliance mechanisms that help users fully leverage the benefits of AI-generated ideas, beyond simply improving AI's output quality.

In the \textit{Test} phase, the pattern shifts. The Kruskal-Wallis H test reveals significant differences in originality across conditions (\textit{H}(2) = 9.14, \textit{p} = 0.010). Post-hoc pairwise comparisons using Dunn’s test with Bonferroni correction indicate that participants exposed to the \textit{List of Strategies} performed significantly worse than those with no LLM exposure (\textit{p} = 0.009). Additionally, the originality of participants in the \textit{List of Strategies} condition decreased from the exposure to the test round, suggesting that they did not successfully internalize the strategies well enough to apply them independently. Neither the comparison between \textit{List of Ideas} and \textit{List of Strategies} (\textit{p} = 0.126), nor the comparison between \textit{List of Ideas} and \textit{No LLM Response} (\textit{p} = 1.000) showed significant differences. This suggests that the \textit{List of Ideas} condition may fall somewhere in between, without a strong directional impact on originality. Overall, these results suggest that participants tended to perform better, in terms of originality, when they had no prior exposure to LLMs in the test phase.

\begin{figure*}
\centering
\includegraphics[width=0.75\textwidth]{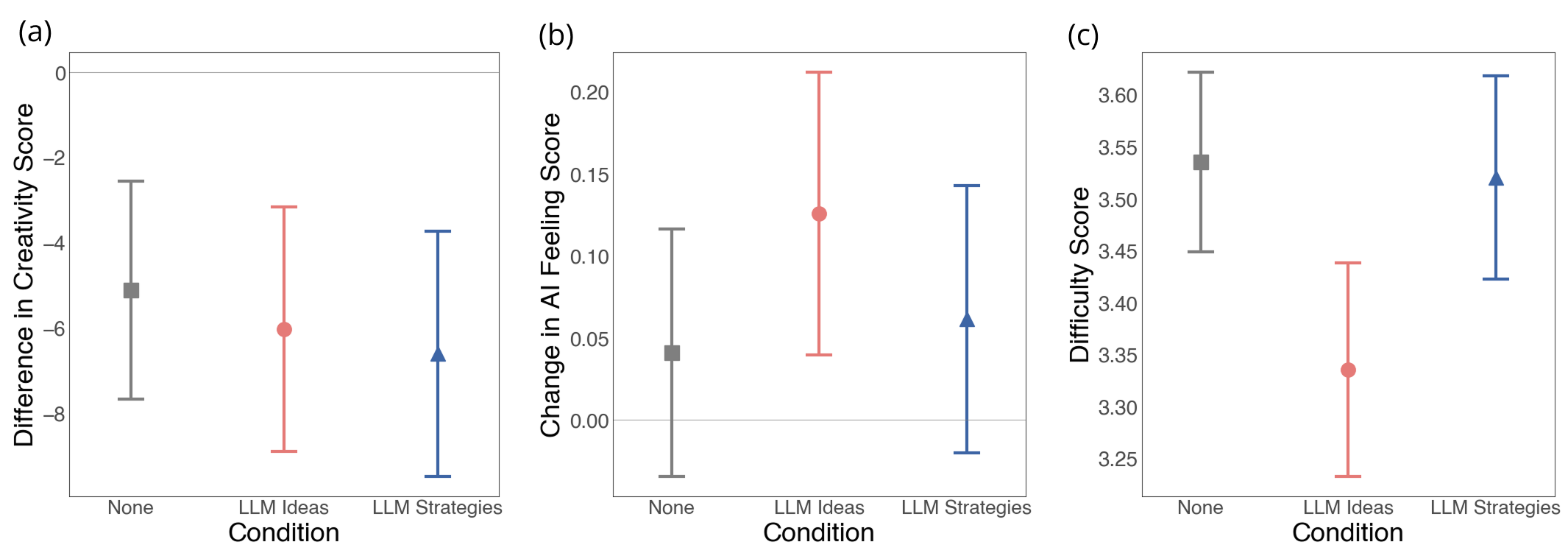}
\caption{Plots of subjective measures collected before and after participants completed the Alternate Uses Task. The left figure shows participants' change in self perceived creativity levels (Based on how many \% of humans they felt they were more creative than), the middle figure indicates how their feelings towards the increased use of AI computer programs in daily life changed (Between More concerned than excited/More excited than concerned/Equally excited and concerned), and the right figure presents how much difficulty they had in coming up with ideas for the test object, all segmented by the three LLM response types.}
\Description{Figure 6 shows plots of subjective measures collected before and after participants completed the Alternate Uses Task all segmented by the three LLM response types. The left figure shows participants’ change in self perceived creativity levels (Based on how many \% of humans they felt they were more creative than) from -8 to 0, No LLM Response at -5, LLM Answer at -6, and LLM Guidance at -7, all with error bars roughly 5 units long. The middle figure indicates how their feelings towards the increased use of AI computer programs in daily life changed from 0 to 0.2, No LLM Response at 0.04, LLM Answer at 0.12, and LLM Guidance at 0.06, all with error bars roughly 0.16 units long. The right figure presents how much difficulty they had in coming up with ideas for the test object from 3.25 to 3.6, with No LLM Response at 3.53, LLM Answer at 3.33, and LLM Guidance at 3.51, all with error bars roughly 0.17 units long.}
\label{fig:divergent-subjective}
\end{figure*}

\subsubsection{Fluency} 
Figure \ref{fig:divergent-results-1}b shows the average number of ideas generated by participants across different conditions. During the \textit{Exposure} phase, the Kruskal-Wallis H test revealed significant differences in fluency across conditions (\textit{H}(2) = 8.57, \textit{p} = 0.014). Participants who received the \textit{List of Strategies} produced significantly fewer ideas compared to those without LLM exposure, as indicated by Dunn’s post-hoc test (\textit{p} = 0.011). This may be because reading and applying strategies require more time, potentially reducing the number of ideas generated, even though the originality of these ideas remained comparable to the \textit{No LLM Response} condition (as discussed earlier). Interestingly, participants who received the \textit{List of Ideas} submitted nearly two fewer ideas than what was shown to them on average per round, being shown seven ideas each round, possibly supporting the hypothesis that individuals may prioritize generating their own ideas over simply adopting AI-provided ones.

In the \textit{Test} phase, however, the Kruskal-Wallis H test did not indicate any significant differences in fluency across conditions (\textit{H}(2) = 1.14, \textit{p} = 0.566), suggesting that prior exposure to LLMs may not significantly impact the quantity of ideas participants can produce independently in subsequent rounds of AUT.

\subsubsection{Creative Flexibility} 
Figure \ref{fig:divergent-results-1}c shows the average creative flexibility across conditions, which measures how different the ideas produced in the Test round are from those produced in the Exposure rounds (a higher value indicates more dissimilarity). The Kruskal-Wallis H test revealed a significant difference in creative flexibility across conditions (\textit{H}(2) = 6.31, \textit{p} = 0.043). However, post-hoc pairwise comparisons using Dunn’s test with Bonferroni correction did not reach significance after correcting for multiple comparisons. Directionally, participants in the \textit{List of Strategies} condition tended to produce ideas in the Test round that were more similar to those from the Exposure rounds compared to other conditions (\textit{p} = 0.052 when compared to \textit{No LLM Response}). Applying the same strategies to different objects may have led to more similar final outcomes across rounds.

\begin{figure*}
\centering
\includegraphics[width=\textwidth]{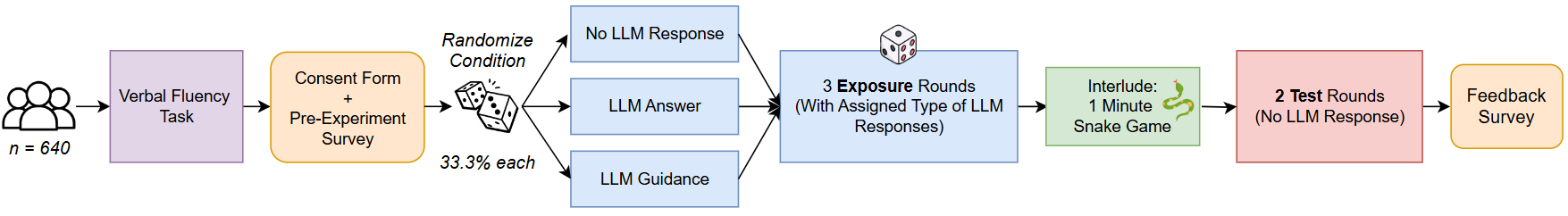}
\caption{Schematic of design for Experiment 2 on convergent thinking.}
\Description{Figure 7 illustrates the high level structural overview of the convergent thinking experiment. 640 participants start with a verbal fluency task, consent form, and pre-experiment survey before being randomized into one of three conditions for three exposure rounds. These conditions are: A (human-only), B (AI provides potential answers), and C (AI provides guidance). After the exposure rounds, participants engage in a 1 minute snake game as a brief interlude before moving onto 2 test rounds that all participants carry out without any AI support. Finally, they end with a feedback survey.}
\label{fig:exp-design-2}
\end{figure*}

\subsubsection{Individual- and Group-level Diversity of Idea Set}
Figure \ref{fig:divergent-results-2}a shows the trend in the individual-level diversity of ideas produced by participants. In the \textit{Exposure} phase, the Kruskal-Wallis H test revealed a significant difference between conditions (\textit{H}(2) = 9.95, \textit{p} = 0.007). Post-hoc pairwise comparisons indicated that participants who received the \textit{List of Strategies} produced significantly more similar ideas within each round compared to those in the \textit{List of Ideas} condition (\textit{p} = 0.009) and the \textit{No LLM Response} condition (\textit{p} = 0.041). In contrast, there was no significant difference between the \textit{List of Ideas} and \textit{No LLM Response} conditions (\textit{p} = 1.000). The Kruskal-Wallis test for the \textit{Test} phase also showed a significant difference between conditions (\textit{H}(2) = 7.71, \textit{p} = 0.021). Participants in the \textit{List of Strategies} condition continued to generate more similar ideas compared to those in the \textit{No LLM Response} condition (\textit{p} = 0.017), although no significant difference was found between the \textit{List of Strategies} and \textit{List of Ideas} conditions (\textit{p} = 0.348), or between \textit{No LLM Response} and \textit{List of Ideas} (\textit{p} = 0.747).

Interestingly, while both LLM-assisted conditions showed a decline in idea diversity across rounds, participants in the \textit{No LLM Response} condition maintained or even slightly improved their diversity. This result is somewhat unexpected, as it might have been assumed that unassisted participants would experience fatigue, leading to less varied ideas over time, whereas those who had support during the exposure rounds would be better equipped to produce diverse ideas.

We simulated group-level diversity by randomly sampling ideas across participants based on their condition, item, and phase. Figure \ref{fig:divergent-results-2}b illustrates the trends in group-level diversity. In the \textit{Exposure} rounds, the average median cosine distance between ideas was similar across all conditions (\textit{p} > 0.05), which is unexpected. Participants in the \textit{List of Ideas} condition, who had access to the same LLM-generated ideas, were expected to produce more similar ideas as a group. However, this may be due to participants not fully adopting the AI suggestions, as indicated by earlier findings. Additionally, the random sampling of 7 ideas from a pool of 20 for each object may have contributed to the lack of overlap.

Interestingly, participants exposed to LLM-generated strategies during the exposure phase continued to generate more similar ideas in the test phase, even without LLM assistance, a phenomenon known as homogenization. This raises concerns that LLMs, which provide ubiquitous frameworks for thinking, could lead to reduced diversity in collective thinking, with people continuing to produce similar ideas even after they stop using the LLM.

\subsubsection{Subjective Measures and Perceptions} 
Figure \ref{fig:divergent-subjective} summarizes participants' self-reported measures of creativity, feelings toward AI use, and the difficulty of generating uses for the \textit{Test} object. Across all conditions, participants reported a decline in their perceived creativity after the experiment, a consistent drop that is somewhat unexpected. One might assume that receiving a list of AI-generated ideas would either further diminish creativity, as participants might feel they cannot match the AI, or conversely, seeing strategies during exposure rounds could boost their self-assessed creativity by providing structured guidance.

The change in participants' feelings toward AI remained relatively stable across conditions, with those in the \textit{List of Ideas} condition reporting more than double the magnitude of change compared to the other groups. Interestingly, participants who received a list of ideas during the exposure rounds also found it easier to generate their own ideas during the test phase. This is surprising, as one might expect that exposure to AI-generated ideas would make it harder for participants to think independently in the test round, and that AI strategies in exposure would make it easier to independently come up with ideas.

\section{Experiment 2: Convergent Thinking}
The results from our first experiment highlight how LLMs shape our ability to generate ideas independently. However, creativity involves more than idea generation; it requires the capacity to identify and refine the most appropriate solutions within a given conceptual space. Convergent thinking is crucial for advancing from a broad set of ideas to selecting the most viable outcome. Motivated by this, our second pre-registered\footnote{\url{https://aspredicted.org/4q75-bskd.pdf}} experiment focuses on convergent thinking, testing not only LLM-generated answers but also the potential of LLMs to guide users toward solutions, simulating a coaching approach.

\subsection{Experimental Design}

We conducted a second three-condition, between-subjects experiment to understand how participants’ unassisted convergent thinking performance are affected by the presence of LLM assistance during prior convergent thinking tasks (see Figure \ref{fig:exp-design-2}). We employed the Remote Associates Test (RAT), a widely recognized task for measuring convergent thinking. For each RAT problem, participants are presented with three words and asked to generate a fourth word that connects or fits all three words within a one-minute timeframe. For example, given the words \textit{shelf}, \textit{log}, and \textit{worm}, the correct response would be `book' (\textbf{book}shelf, log\textbf{book}, \textbf{book}worm). We selected this task due to its widespread use in prior research \cite{koutstaal2022curiosity, cortes2019re},  and its compatibility with LLMs. 

Similarly to the previous experiment, this study included both exposure (paired) and test (solo) phases. Figure \ref{fig:exp-design-2} gives a high-level overview of the experiment schema. The experiment had participants complete a series of RAT problems alongside AI assistance (exposure phase), before completing unassisted RAT problems (test phase). The experiment was designed to give us insight into how convergent thinking skills are impacted by different levels of AI assistant, as well as by immediate prior use of AI assistance. Because a given RAT problem has a singular correct answer, we were able to define the metric for convergent thinking skills as accuracy on the RAT problems. We additionally collected perceptions and sentiments before and after the experiment to see how these were impact by completing the tasks alongside AI assistance or not.

\begin{figure*}
\centering
\includegraphics[width=\textwidth]{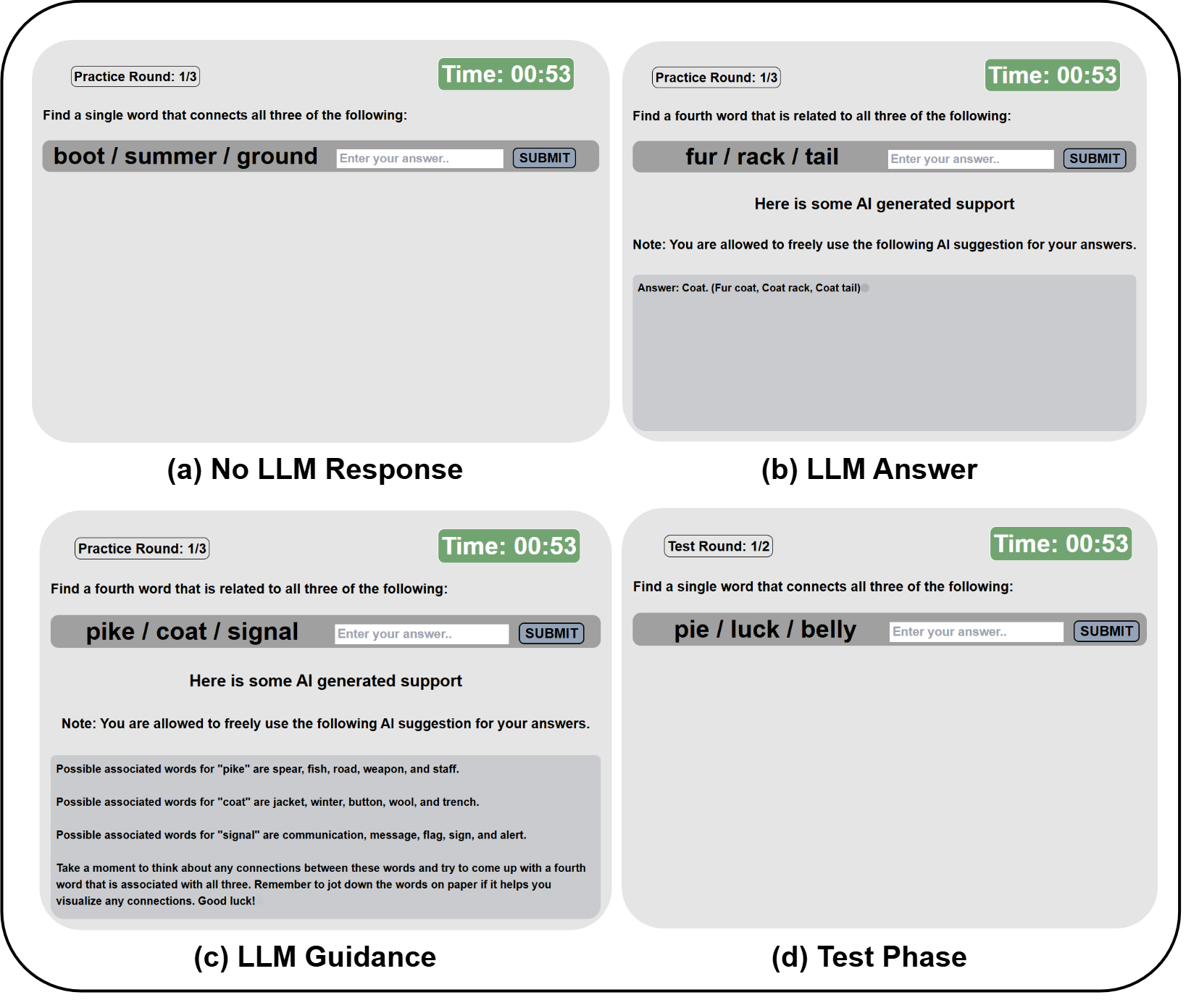}
\caption{Interface used in the convergent thinking experiment across all 3 exposure round conditions and test rounds.}
\Description{Figure 8 shows the 4 user interfaces used in the convergent thinking experiment, (No LLM Response, LLM Answer, and LLM Guidance exposure conditions, as well as for the test round). The No LLM Response and Test round interfaces are the exact same. A display showing the current round at the top left, and a timer at the top right. Right below this is some brief instruction, and right below that is space for participants to enter their answer. The LLM Answer and LLM Guidance interfaces have an extra section situated at the bottom on top of all the above. That is where the text containing the LLM assistance is being displayed.}
\label{fig:exp-screens-2}
\end{figure*}

\subsubsection{Conditions and treatment} 

 During the exposure phase, participants were required to solve three RAT problems consecutively. They were randomly assigned to one of three conditions, each providing different types of LLM assistance:

\begin{tabular}{@{}p{2.2cm}@{}p{5.8cm}@{}}
    \textit{None:} & Control group. \\
    \textit{LLM Answer:} & The answer generated by GPT-4o for the given set of words. \\
    \textit{LLM Guidance:} & GPT-4o was customized with a system prompt (shown in Figure \ref{fig:pre-prompt2}) to generate possible associated words for each of the three given words. The model was instructed not to provide the solution directly but to encourage participants to make connections on their own, for example, by jotting the words down on paper. \\
\end{tabular}

Similar to Experiment-1, the \textit{LLM Guidance} condition was inspired by a common real-world method of using LLMs to obtain structured thinking frameworks rather than direct answers. After the exposure phase, participants played a game of Snake for one minute which served as a distractor task. In the test phase, all participants completed two additional RAT rounds (randomly selected) to attempt without any LLM assistance.

The \textit{No LLM Response} (Figure \ref{fig:exp-screens-2}a) condition serves as a baseline, reflecting the state of convergent thinking without AI assistance. The \textit{LLM Answer} (Figure \ref{fig:exp-screens-2}b) condition represents the scenario in which AI provides ready-made solutions. The \textit{LLM Guidance} (Figure \ref{fig:exp-screens-2}c) condition was designed to assist participants in navigating the conceptual space of the problem without directly revealing the answer. This approach aims to enable participants to solve the problem independently while still offering useful guidance in the moment. Research has shown that priming with related words and manually writing responses can enhance performance on RATs---strategies a coach might use for guidance \cite{carlsson2019effects, mednick1964incubation}.

In all conditions, the assigned type of LLM response appeared 5 seconds after the question was shown to the participant, with the text being displayed character by character, mimicking typical chat-based LLM interfaces. In the \textit{LLM Answer} and \textit{LLM Guidance} conditions, the interface informed participants they could freely use the AI suggestions in their answers. Although the RAT has participants employ creative thinking, there is always one definitive answer. As such, we wanted to avoid the case in which a participant was prompted with the correct answer by AI, acknowledged that it was the correct answer, but then felt as though they should attempt to find a different solution to be ``more creative''.

During the exposure rounds, participants were made to wait the full 1 minute before advancing to the next task, while in the test rounds, participants could advance once they submitted an answer. This was to discourage participants from speeding through the experiment without giving thought to their answers. Like Experiment-1, we control for time per question to map onto real-world scenarios of LLM use in work environments. Participants were never shown the correct answer to the RATs they completed (except as part of LLM responses, if applicable), nor were they informed if their answers were correct.

\begin{figure}
\centering
\includegraphics[width=0.4\textwidth]{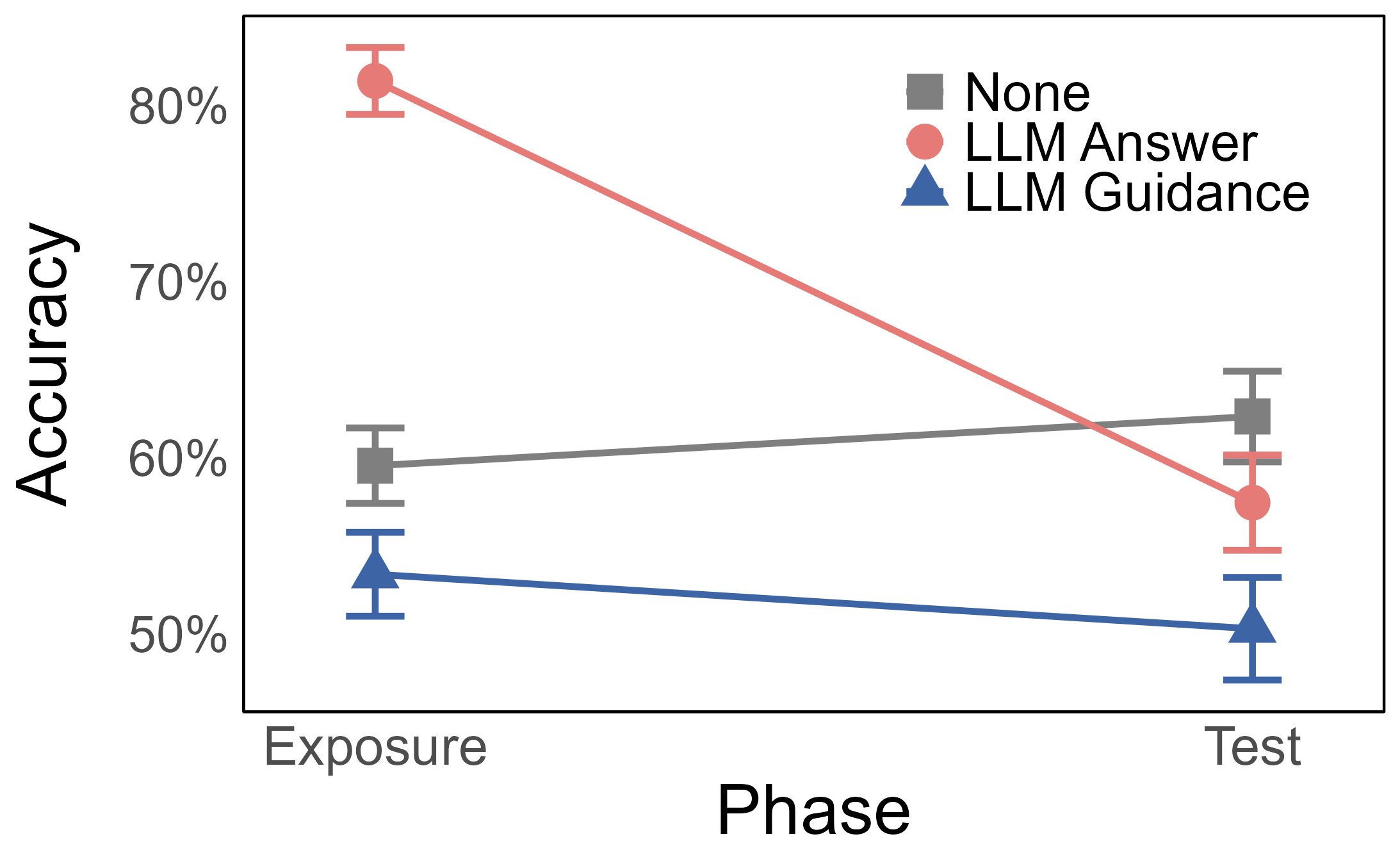}
\caption{Plot of participant accuracy on the Remote Associates Task, segmented by exposure and test phase across all experimental conditions.}
\Description{Figure 9 shows Accuracy(\%) from 50 to 80 on the Y axis, and phase (exposure and test) on the X axis. No LLM response shows a slight increase from 60 to 62, LLM Answer shows a decrease from 80 to 57 and LLM Guidance shows a slight decrease from 54 to 50.}
\label{fig:convergent-accuracy}
\end{figure}

\subsubsection{Stimulus} 
For the Remote Associates Test (RAT), we employed a widely-used dataset from Bowden and Jung-Beeman~\cite{bowden2003normative}. The subset we selected contained 45 questions equally distributed across three difficulty levels: easy, medium, and hard. 
For our study, we randomly selected questions from the easy and medium difficulty levels for each round, as pilot testing indicated that these were appropriate for our participant pool (crowdworkers). Figure~\ref{fig:rat-responses} provides examples of the LLM responses presented to participants in the different experimental conditions. 

\subsubsection{Verbal Fluency and Surveys}
At the start of the experiment, participants completed a verbal fluency task, where they were asked to list as many English words as possible within one minute, starting with a given letter (e.g., ‘F’). We implemented this measure to account for participants with low English language proficiency, which can significantly confound performance on the RAT~\cite{hedden2005fluency, wang2016proficiency}. Following this, participants received instructions for the RAT, along with additional information specific to each experimental condition. After the instructions, participants completed a survey assessing: 

\begin{itemize}
  \item \textbf{Self-reported creativity}, measured as a sliding value from 1-100, with the prompt: I am more creative than X\% of humans.
  \item \textbf{Attitudes toward AI use in daily life}, measured as a multiple choice question with options; “More concerned than excited”, “Equally excited and concerned”, and “More excited than concerned”.
\end{itemize}
After finishing the final test round, participants were surveyed again on these same questions, as well as:
\begin{itemize}
  \item \textbf{Perceived test round difficulty}, measured with the prompt “How difficult was it to come up the associated word for the last two (test) tasks?” with options; “Very easy”, “Somewhat easy”, “Somewhat difficult”, and “Very difficult”.
  \item \textbf{Perceived helpfulness of exposure rounds} measured with the prompt “How helpful was the exposure phase (first three questions)?” with options; “Not at all helpful”, “A little helpful”, and “Very helpful”
\end{itemize}
The post-experiment survey also contained a basic attention check.

\subsubsection{Analysis} Following our pre-registration, we conducted an Analysis of Covariance (ANCOVA) to compare the average accuracy between the three conditions in test rounds, controlling for the number of words generated in the verbal fluency task as a covariate. We used Tukey's Honestly Significant Difference (HSD) test as a post-hoc test for pairwise comparisons between conditions after a significant ANCOVA result.

\subsubsection{Participants}
We recruited 640 participants from Prolific (of whom 99\% passed the attention check). The sample size was determined through a power analysis using simulated data from pilot studies, ensuring 80\% power for the main pre-registered test. Participants were based in the US or UK, and were fluent in English. The task took approximately 10 minutes to complete and the participants were paid \$1.30. On average, the participants felt they were more creative than 46.4\% of the population, and they were able to generate 13.3 words in the verbal fluency task. 

\begin{figure*}
\centering
\includegraphics[width=0.8\textwidth]{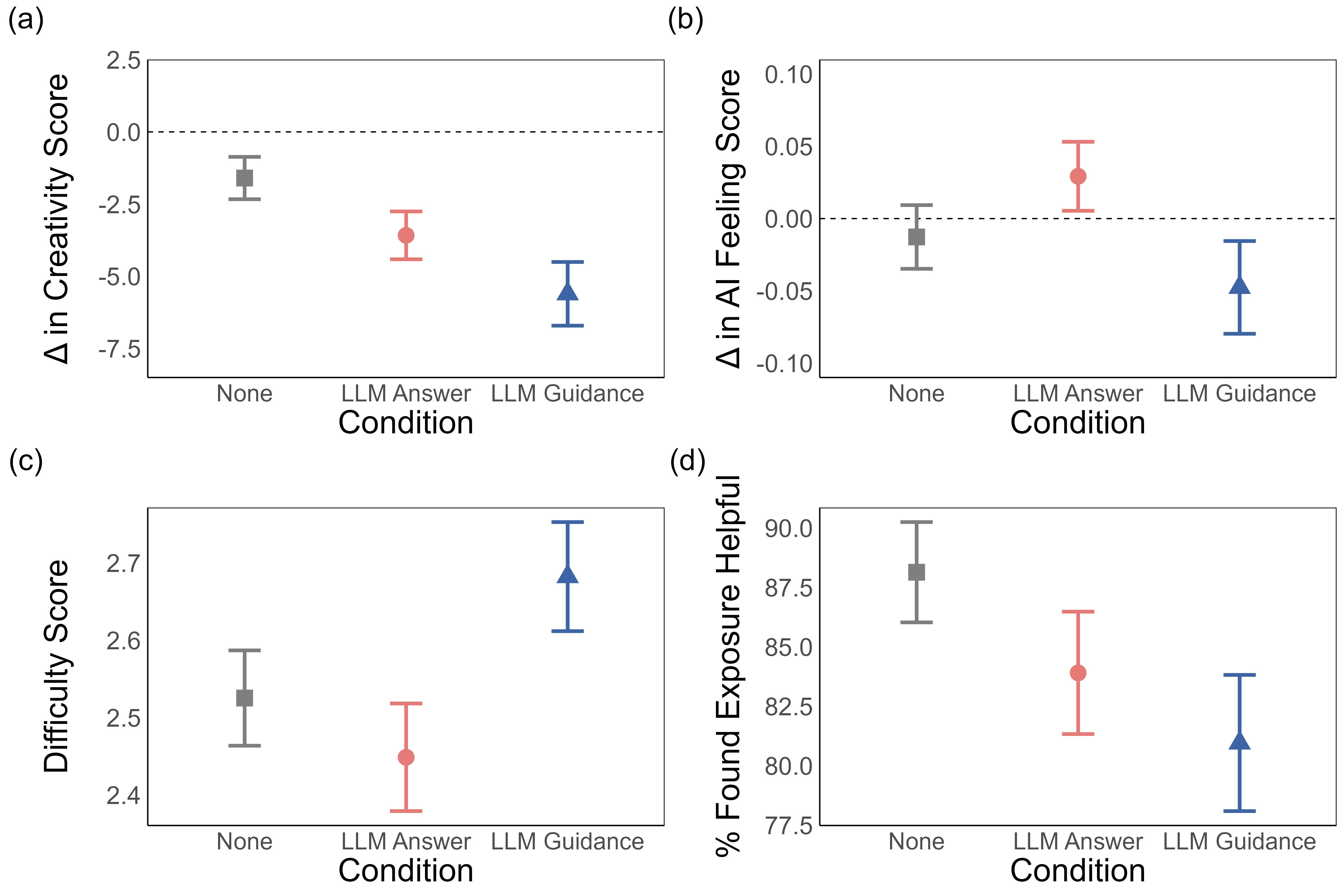}
\caption{Plots of participant accuracy on the Remote Associates Task, segmented by exposure and test phases across all experimental conditions. The top left figure shows participants’ change in self perceived creativity ratings (Based on how many \% of humans they felt they were more creative than), the top right figure indicates how their feelings towards the increased use of AI computer programs in daily life changed (Between More concerned than excited/More excited than concerned/Equally excited and concerned), the bottom left figure displays how much
difficulty they had in coming up with answers for the test phase, and the bottom right figure presents how many participants found the exposure rounds helpful.}
\Description{Figure 10 shows 4 plots detailing participant perceptions across conditions, None, LLM Answer and LLM Guidance. The top left plot shows Change in Reported Self-Creativity Rating from -7.5 to 2.5 on the Y axis and condition on the X axis. No LLM Response is at -2, LLM Answer is at –3.5 and LLM Guidance is at -6. The top right plot shows Change in Feeling about AI Use from -0.10 to 0.10 on the Y axis and condition on the X axis. No LLM response is at -0.02 LLM Answer is 0.025 and LLM Guidance is -0.05. The bottom right plot shows Perceived Difficulty of Test Rounds from 2.4 to 2.7 on the Y axis and Condition on the X axis. No LLM response is at 2.52, LLM Answer is at 2.45 and LLM Guidance is at 2.66. The bottom right plot shows Percentage of Participants Who Found Exposure Rounds Helpful from 77.5 to 90 on the Y axis, and condition on the X axis. No LLM response is at 88 LLM Answer is at 83.5, and LLM Guidance is at 81.}
\label{fig:convergent-survey}
\end{figure*}

\subsection{Results}
\subsubsection{Accuracy in the Test Rounds}

As shown in Figure \ref{fig:convergent-accuracy}, participants in the \textit{LLM Answer} condition did better in the Exposure rounds in comparison to the other conditions. However, this trend changed in the Test rounds. Following our pre-registered analysis plan, the ANCOVA revealed a significant effect of verbal fluency on performance (\(F(1, 1256) = 16.715, p < 0.001\)) as well as a significant main effect of condition (\(F(2, 1256) = 5.134, p = 0.006\)). Post-hoc comparisons using Tukey’s HSD test indicated that participants in the \textit{LLM Guidance} condition performed significantly worse than those who did not receive any LLM responses in exposure rounds (\(p = 0.005\)), with a mean difference of \(-10.6\) [95\% CI: \(-18.6, -2.7\)]. While the \textit{LLM Answer} condition did not differ significantly from the \textit{No LLM Assistance} condition (\(p = 0.682\)), there was a trend suggesting that the \textit{LLM Guidance} condition performed worse than the \textit{LLM Answer} condition (\(p = 0.064\)).

This suggests that, although the difference between the \textit{LLM Answer} and \textit{LLM Guidance} conditions did not reach statistical significance levels, the \textit{LLM Answer} condition may lie somewhere in between the other two conditions. However, this observation should be interpreted cautiously, given the lack of statistical significance in this comparison.

The positive performance of participants in the \textit{LLM Answer} condition during exposure rounds underscores the effectiveness of LLMs in providing accurate solutions to the task. This suggests that LLMs are indeed proficient at generating correct answers and that participants are capable of recognizing when to leverage LLM advice effectively. Exposure to LLM assistance—whether in the form of direct answers or strategic guidance—did not translate into enhanced unaided performance and may have even been counterproductive. This phenomenon can be understood through the lens of convergent thinking, which often relies on achieving an 'aha' moment or insight. \textit{LLM Guidance} might have disrupted this process by providing too much additional information to take in, thereby hindering the natural cognitive processes necessary for independent problem-solving. Similarly, \textit{LLM Answer} could have impeded participants' engagement in their own creative thinking, as they might have become overly reliant on external solutions rather than developing their own problem-solving strategies.

\subsubsection{Subjective Measures and Perceptions}

Figure \ref{fig:convergent-survey} shows participants' change in self-reported levels of creativity, change in feelings towards AI use perceived difficulty of test rounds, and perceived helpfulness of exposure rounds. In \textbf{plot (a)}, we see that across all experimental conditions, participants exhibited a reduction in self-reported creativity ratings from pre-experiment to post-experiment surveys. Both \textit{LLM Answer} and \textit{LLM Guidance} conditions saw a greater average reduction in creativity ratings, with \textit{LLM Guidance} showing an average reduction more than twice as substantial as that observed in the \textit{No LLM Assistance} condition. This suggests that reliance on LLMs, particularly in the \textit{LLM Guidance} condition, may have undermined participants' confidence in their own creative abilities. The more substantial reduction in creativity ratings for \textit{LLM Guidance} could indicate that receiving a wide range of possible word connections, rather than necessarily coming up with their own, diminished participants' sense of ownership over their creative process. Futrthermore, the less significant decrease observed in the \textit{LLM Answer} condition could be from participants recognizing they were receiving direct answers which were completely detached from their own creative abilities. \textbf{Plot (b)} shows that the change in attitudes toward AI use in daily life stayed roughly constant between pre-experiment to post-experiment surveys. with \textit{LLM Answer} seeing slight attitude increase. Feelings towards AI use was measured using a 3-point Likert scale (with values being ``More concerned than excited'', ``Equally excited and concerned'' ``More excited than concerned''). This slight increase in positive attitudes toward AI in the \textit{LLM Answer} condition could be attributed to participants recognizing the effectiveness of LLMs in completing tasks accurately, reinforcing their confidence in AI's capabilities. The reliable performance of the LLM may have led to a more favorable view of its potential for everyday use.
In \textbf{plot (c)}, perceived difficulty of the test rounds was highest for the \textit{LLM Guidance} condition, and slightly lower in the \textit{No LLM Assistance} and \textit{LLM Answer} conditions. That is to say, test rounds felt more difficult when the participant had completed previous tasks with guiding AI assistance, rather than just a straight answer, or no assistance at all. On average, perceived difficulty was slightly above 2.5 on a 4-point Likert scale (ranging from ``very easy'' to ``very difficult''). meaning that participants found test rounds a little on the difficult side.
The higher perceived difficulty in the \textit{LLM Guidance} condition may be due to participants relying on the given word associations during the exposure rounds, which might have made it harder for them to independently generate associations in the test rounds where no assistance was given. Finally, \textbf{plot (d)} measured the percentage of participants who found the exposure helpful for their completion of the test rounds. Note that this was not explicitly measuring whether participants found the LLM assistance helpful, but rather whether completing the exposure rounds aided them in performing the test rounds. Participants who received no LLM assistance during the exposure rounds were the most likely to find these rounds helpful. In comparison, the proportion of participants who found the exposure rounds helpful was roughly 5\% lower in the \textit{LLM Answer} condition and 7\% lower in the \textit{LLM Guidance} condition. This result may suggest that participants without LLM assistance had to rely more on their own problem-solving strategies during the exposure rounds, which could have enhanced their perceived usefulness of these rounds in preparing for the test phase. In contrast, those receiving LLM assistance might have become more dependent on the AI, leading to a lower perception of the exposure rounds' value for their independent performance during the test rounds.

\section{Discussion}

\subsection{Key Findings}
\subsubsection{LLMs Boost Performance During Exposure, but Unassisted Participants Excel in Test}
Across both experiments, we observed that LLM assistance was helpful during the exposure phase, aligning with previous research that highlights LLMs' ability to enhance creative task performance when users have access to AI-generated ideas or strategies \cite{anderson2024homogenization, chakrabarty2024art}. However, in the test phase, participants who had no prior exposure to LLMs consistently performed better (this was not statistically significant in all cases). In the divergent thinking task, participants who worked without LLM assistance generated more original ideas on average, and in the convergent thinking task, those without LLM exposure were better able to identify the correct connecting word compared to those who had LLM exposure. These findings suggest that while LLMs may provide short-term boosts in creativity during assisted tasks, they might inadvertently hinder independent creative performance when users are asked to perform without assistance. This raises important questions about the long-term impact of repeated LLM use on human creativity and cognition.

From a design perspective, it is critical to consider not just human-AI performance during exposure phases but also human performance in unassisted tasks after using AI. Systems should be designed with long-term human flourishing in mind, ensuring that the benefits of AI assistance do not come at the cost of diminished independent creative abilities. Ensuring that users can effectively transition from AI-supported creativity to autonomous creative work will help mitigate potential long-term harms associated with over-reliance on LLMs, as highlighted by concerns in the literature about cognitive decline with repeated AI use \cite{heersmink2024use}.

\subsubsection{Differential Impact of LLMs on Divergent and Convergent Thinking}
Our experiments reveal that the effects of LLMs vary significantly depending on the aspect of creativity being measured. In divergent thinking tasks, where participants were asked to generate a wide range of ideas, we observed more skepticism toward LLM assistance. Participants seemed less inclined to adopt AI-generated suggestions, which may be due to the nature of divergent thinking itself—encouraging exploration and unconventional approaches \cite{runco1991divergent, runco2012divergent, baer2014creativity, guilford1967creativity}. In contrast, for the convergent thinking task, where participants were tasked with narrowing down the ideas to a single solution, LLM assistance during the exposure phase appeared more beneficial. This is consistent with the idea that convergent thinking is more structured and goal-oriented, making it easier for participants to recognize when LLMs are effectively guiding them toward the correct solution \cite{mednick1968remote, cropley2006praise, simonton2015praising}.

LLMs, however, may also hinder the creative process. In divergent thinking, the introduction of AI-generated ideas may distract participants, consume valuable cognitive resources, or prevent full engagement with the task. This aligns with theories of creativity such as Boden's concept of \textit{conceptual spaces}, which emphasize the importance of users exploring and navigating creative possibilities independently \cite{boden2010creativity}. Over-reliance on LLMs during divergent thinking may disrupt this exploration, leading to less original ideas, as seen in the originality results (Figure \ref{fig:divergent-results-1}). Similarly, in convergent thinking tasks, LLMs might steer participants toward specific solutions too quickly, reducing the need for deep engagement with the problem space \cite{guilford1967creativity}.

These findings underscore the importance of carefully calibrating trust and reliance in human-LLM systems. Designers should incorporate measures that help users recognize when to trust and rely on LLMs, and when to prioritize their own cognitive processes. This approach can help balance the benefits of AI support with the risks of undermining the human creative process, ensuring that LLMs enhance rather than detract from creativity in the long term.

\subsubsection{Persistent Homogenization of Ideas and the Challenges of Designing LLM Coaches}
Existing research has shown that LLM usage can lead to the homogenization of ideas within groups, where participants tend to converge on similar outcomes when using AI-generated suggestions \cite{anderson2024homogenization, padmakumar2023does, doshi2024generative}. Our findings extend this concern, showing that even when people stop using LLMs that provide strategic frameworks for thinking, the homogenization effect can persist. In our divergent thinking experiment, participants who received LLM-generated strategies exhibited reduced diversity in their idea sets both during and after LLM use, suggesting that such frameworks may have a lasting impact on creative processes, potentially stifling the generation of more varied or unconventional ideas.

Interestingly, we did not find the same effect for participants who received direct ideas from a standard LLM. The \textit{List of Ideas} condition did not lead to the same lasting homogenization once the LLM was no longer present. This indicates that while LLM-generated strategies can have long-term effects on creative diversity, providing ideas without a guiding framework may allow for more cognitive flexibility once participants stop using AI. Further, participants with no LLM interactions in the study collectively generated the most diverse idea sets (Figure \ref{fig:divergent-results-2}). In our convergent thinking experiment, participants who received LLM guidance during exposure performed worse in the test round compared to those who received direct answers. This highlights the complexity of designing LLMs as coaches or guides. While direct answers may not always seem ideal, in some cases, they can be more effective than offering frameworks or strategies. Designers of Human-AI systems must take these findings into account, ensuring that LLM interactions are structured to avoid unintentional long-term effects on cognitive diversity. Controlled experiments, such as those conducted here, can be useful methods in refining and optimizing the design of coach-like LLM systems \cite{spatharioti2023comparing, kumar2023math}.

\subsection{Broader Implications for Fields Involving Human-AI Co-Creativity}
When designing AI tools for co-creativity, it is crucial to consider their long-term impact on human cognitive abilities. Hofman \textit{et al.} \cite{hofman2023steroids} introduced a useful metaphor—steroids, sneakers, and coach—to describe the spectrum of AI’s role in human-AI collaboration. Our findings suggest that co-creative systems must be carefully designed to be coach-like to prevent unintended consequences, such as stifling human creativity, even after AI assistance is removed.

These insights have broad implications for various fields, from scientific discovery to the arts and humanities. In the context of AI for science, for instance, there is ongoing excitement about building increasingly sophisticated models to accelerate the process of hypothesis generation and experimental execution \cite{boiko2023chemical, si2024llmsgeneratenovelresearch, lu2024aiscientistfullyautomated}. However, the design of these models often overlooks their potential impact on scientists’ creative abilities. Although some initial explorations have been conducted, the field lacks rigorous empirical evaluation of how AI systems affect human creativity in scientific discovery. The arts and humanities may face similar challenges \cite{rane2024}. For example, writers using the same LLM could produce homogenized content, even when AI is no longer part of their workflow. This underscores the need for AI systems that not only assist in the creative process but also promote long-term cognitive diversity, ensuring that human creativity thrives in collaboration with AI rather than becoming constrained by it. 

Our work in a controlled, simple setting suggests that the performance of AI models alone is not enough to determine their value. The design of human-AI interactions—whether AI provides direct answers or encourages users to think critically—and the timing of performance evaluation (during AI use versus after) both significantly alter the narrative. These considerations are critical in ensuring that AI models enhance, rather than diminish, human creative potential. Beyond simply building superhuman AI, we must focus on how these tools influence human creativity, culture, and cognitive growth, aiming for AI systems that enrich and elevate human thought \cite{brinkmann2023machine, heersmink2024use}.

\subsection{Limitations \& Future Work}
While our study provides valuable insights into the impact of LLMs on human creativity, there are several limitations that warrant further investigation.

Firstly, measuring creativity itself presents challenges. Though we employed both divergent and convergent thinking tasks (AUT and RAT) to capture a broad spectrum of creative processes, these tasks were limited to verbal responses and conducted over short time periods. Creativity, however, is a much more complex and nuanced phenomenon. Future work should expand the battery of tasks to include more natural, real-world creative activities, such as writing advertisements, designing magic tricks, or solving complex problems. Additionally, creativity tasks that are non-verbal or visual, like those supported by diffusion models, could be explored. Experiments using visual creativity tasks, such as the Test of Creative Thinking Drawing Production (TCT-DP) or the Evaluation of Potential Creativity (EoPC) \cite{lubart2016convergent}, would provide deeper insights into how LLMs impact creativity in non-text domains.

Our study also focused exclusively on the active, conscious aspects of creativity, largely due to the controlled lab environment. Yet, creativity often involves unconscious processes that are harder to capture in such settings. Another limitation lies in the short exposure period in our lab-based study. In real-world settings, exposure to LLMs is often prolonged, integrated into daily workflows, and thus, likely produces more significant and lasting effects on creativity. In contrast, our study's shorter exposure periods may have resulted in smaller effect sizes, limiting our ability to fully understand the long-term implications of LLM use. Future research should explore prolonged interactions with LLMs to better reflect real-world scenarios and study potential cumulative impacts on creativity. We acknowledge that limiting the AUT objects in Experiment-1 to five is a limitation. We selected them due to the strong correlation between our automated scoring method and human judgments for these objects, which enabled us to evaluate ~10,000 ideas at scale without prohibitive human annotation, in line with other large-scale creativity studies that have operated under similar constraints \cite{ashkinaze2024ai}.

Additionally, the design of our LLM-based guidance presents limitations. In the convergent thinking task, the guidance provided was naturally more verbose than direct answers, which may have inadvertently influenced the results. The list-of-strategies and guidance conditions introduced more cognitive load and hence the comparisons with the default conditions during exposure may have been unfair. Future studies should control for verbosity and ensure that the nature of guidance is standardized to more accurately assess the impact of LLM ``coaching'' versus direct assistance. However, the additional cognitive load during exposure may be beneficial for human cognition long-term as has been shown in education-related experiments. Another approach could be to measure the human performance in the final 30 seconds of the AUT task as the participants may have completed reading/processing LLM output by this time. Future work can also explore the trends in performance split by the different rounds of the experiment. Moreover, participants in the \textit{coach} conditions may have performed worse during the test due to fatigue build-up in the exposure rounds. Or it could be due to the increased dependency on LLM responses. Future work should investigate the exact mechanisms behind these effects. Finally, although this study represents an early attempt to experimentally measure the impact of LLMs on human creativity, the static nature of our LLM interaction may not fully reflect real-world applications. In practical use cases, AI tools often engage in dynamic, interactive exchanges where users can refine inputs, seek clarifications, or adjust outputs. Future research should investigate how conversational and adaptive LLMs influence creativity over time, as these systems more closely resemble how people use AI in real-world creative processes.

\section{Conclusion}
Through this work, we sought to understand the impact of LLMs on human creativity. Our work moves beyond a single-use scenario or model-only evaluations. We conducted two parallel experiments on divergent and convergent thinking, two key components of creative thinking. By exploring repeated exposure, we attempt to approximate real-world interactions with LLMs and investigate the lasting impacts on human creative processes. Taken together, these experiments shed light on the complex relationship between human creativity and LLM assistance, suggesting that while AI can augment creativity, the mode of assistance matters greatly and can shape long-term creative abilities. In closing, we hope this work offers a template to experimentally evaluate the impact of generative AI on human cognition and creativity.

\begin{acks}
We are grateful to Polly Denny for introducing us to philosophical and historical perspectives on creativity. We are grateful to Jake Hofman for his feedback on our experimental design, and to him, Dan Goldstein, and David Rothschild collectively for their sports metaphor on human–AI relationships, which inspired much of this work. We also appreciate Hope Schroeder for recommending pivotal GenAI+creativity papers, and Jessica Bo, Lilio Mok, and members of the CSSLab at the University of Toronto for their valuable discussions leading up to the experiments. Finally, we thank Peter Organisciak and his group for generously providing access to their automated scoring API for AUT ideas.
\end{acks}

\bibliographystyle{ACM-Reference-Format}
\bibliography{references}

\appendix

\section{Experiment 1 (Divergent Thinking)}
\subsection{LLM Configuration}

\subsubsection{Model Specification}
\begin{itemize}
  \item \textbf{model version}: gpt-4o
  \item \textbf{date of use}: August 2024 
  \item \textbf{temperature}: 0
  \item \textbf{max response}: 4000
  \item \textbf{top-p}: 0.95
  \item \textbf{frequency penalty}: 0
  \item \textbf{presence penalty}: 0
\end{itemize}

\subsubsection{Input Prompt:
\textit{"What are some creative uses for a [OBJECT]? The goal is to come up with creative ideas, which are ideas that strike people as clever, unusual, interesting, uncommon, humorous, innovative, or different. List creative uses for a [OBJECT]."}
} 

\subsubsection{System Prompt} 
``List of Strategies'' condition (Figure \ref{fig:pre-prompt1})
\begin{figure}[htbp]
    \centering
    \fbox{
        \begin{minipage}{0.48\textwidth}
            {\small
            You are an AI assistant tasked with the specific goal of teaching users to be more creative by applying the SCAMPER technique. SCAMPER is a powerful tool for creative thinking that involves exploring various ways to modify an object or idea to generate new possibilities. Your interactions should guide users through the SCAMPER process, which includes the following steps:
            \begin{itemize}
                \item \textbf{Substitute:} Encourage users to think about what elements of the object could be replaced with something else to improve or give it a new use.
                \item \textbf{Combine:} Prompt users to consider how they might combine the object with other items or ideas to create something new.
                \item \textbf{Adapt:} Ask users how the object could be adapted or modified to serve another purpose, or how it might be inspired by something else.
                \item \textbf{Modify:} Challenge users to think about how altering the shape, size, color, or any other attribute of the object could create new uses or improvements.
                \item \textbf{Put to another use:} Encourage users to brainstorm completely different uses for the object, beyond its original purpose.
                \item \textbf{Eliminate:} Guide users to consider what might happen if parts of the object were removed or simplified. How does this change its potential uses?
                \item \textbf{Reverse:} Ask users to consider what would happen if they reversed the object's purpose or use. Would it function differently or serve a new role?
            \end{itemize}
            In your guidance, encourage users to iterate through these steps, not necessarily in order, to explore the full range of creative possibilities. Don’t mention SCAMPER explicitly in your response. Output a list that the user can follow. Don't include any examples, but only prompts or questions to help users ideate. Make your response concise and easy to follow for the users. Express each point in as few words as possible.
            }
        \end{minipage}
    }
    \caption{The system prompt for the ``List of Strategies'' condition (not visible to participants).}
    \label{fig:pre-prompt1}
\end{figure}

\subsection{Sample LLM Responses}
Table \ref{tab:aut-responses} shows the sample LLM responses in Experiment-1.
\begin{table*}[htbp]
\centering
\renewcommand{\arraystretch}{1.2}
\setlength{\tabcolsep}{5pt}
\begin{tabular}{|p{0.15\textwidth}|p{0.4\textwidth}|p{0.4\textwidth}|}
\hline
\textbf{AUT Object} & \textbf{LLM Ideas} & \textbf{LLM Strategies} \\
\hline
Tire 
& 
Garden Planter: Cut a tire in half and use it as a unique planter for flowers or vegetables. \newline \newline
Swing: Hang a tire from a tree to create a fun outdoor swing for kids.\newline \newline
Coffee Table: Stack a few tires on top of each other, paint them, and add a glass top to create a unique coffee table.\newline \newline
Sandbox: Fill a large tire with sand to create a small sandbox for children.\newline \newline
Tire Ottoman: Stuff a tire with cushioning material and cover it with fabric to create a unique ottoman.\newline \newline
Tire Pond: Dig a hole in your backyard, place a tire in it, and fill it with water to create a small pond.\newline \newline
Tire Climbing Wall: Attach several tires to a wall to create a climbing wall for kids. 
&
What can you replace on the tire to improve it or give it a new use? \newline \newline
How can you combine the tire with other items or ideas to create something new? \newline \newline
How can the tire be adapted or modified to serve another purpose? \newline \newline
How can altering the shape, size, or color of the tire create new uses or improvements? \newline \newline
What completely different uses can you brainstorm for the tire, beyond its original purpose? \newline \newline
What might happen if parts of the tire were removed or simplified? How does this change its potential uses? \newline \newline
What would happen if you reversed the tire's purpose or use? Would it function differently or serve a new role?"\\
\hline
Pants
& 
Plant Pot Cover: Use pants to cover unsightly plant pots. The pockets can also hold gardening tools. \newline \newline
Art Canvas: Use the fabric of the pants as a canvas for painting or drawing. \newline \newline
Hammock for Small Pets: Sew the ends of the legs together to create a cozy hammock for small pets. \newline \newline
Wine Bottle Bag: Use a pant leg as a chic, recycled wine bottle gift bag. \newline \newline
Patchwork Quilt: Combine various pants to create a unique patchwork quilt. \newline \newline
DIY Bag: Convert an old pair of jeans into a stylish tote bag. \newline \newline
Halloween Decorations: Stuff the pants with hay or newspapers and pair with a shirt to make a scarecrow. \newline \newline
&
What could you replace the fabric or parts of the pants with to create something new? \newline \newline
How can you combine pants with another item to make a unique product? \newline \newline
How could you modify pants to serve a completely different purpose? \newline \newline
What happens if you change the shape, size, or color of the pants? \newline \newline
What entirely different function could pants serve? \newline \newline
What if you remove certain parts of the pants? How does this change their use? \newline \newline
What if you reverse the way pants are used or worn?
What new function could they serve? \\
\hline
\end{tabular}
\caption{Sample LLM responses for the exposure rounds in the divergent thinking experiment.}
\label{tab:aut-responses}
\end{table*}

\section{Experiment 2 (Convergent Thinking)}
\subsection{LLM Configuration}
\subsubsection{Model Specification}
\begin{itemize}
  \item \textbf{model version}: gpt-4o
  \item \textbf{date of use}: August 2024 
  \item \textbf{temperature}: 0
  \item \textbf{max response}: 800
  \item \textbf{top-p}: 0.95
  \item \textbf{frequency penalty}: 0
  \item \textbf{presence penalty}: 0
\end{itemize}
\subsubsection{Input Prompt}
"In this task, You will be shown sets of three words at a time. For each set of three words, you will be asked to generate a fourth word that connects or fits with all three words. Specifically, the word that you generate, when combined with each of the three stimulus words, should create three word-pairs, each of which makes a common compound word or phrase. Please try to be creative and appropriate in your responses.\newline Find a fourth word that is related to all three of the following:\newline
[word1], [word2], [word3]"

\begin{figure}[H]
    \centering
    \fbox{
        \begin{minipage}{0.49\textwidth}
            {\small
            You are a supportive coach guiding participants through solving Remote Associates Test (RAT) problems. For each of the three given words, provide a concise response by listing five possible associated words. Encourage the participant to identify connections between these words to find the fourth word, but do not directly give the solution. Conclude by suggesting they consider if any word connects all three and nudge them to jot the words down on paper if it might help.
            }
        \end{minipage}
    }
    \caption{The system prompt for the \textit{LLM Guidance} condition (not visible to participants).}
    \label{fig:pre-prompt2}
\end{figure}
\subsection{Sample LLM Responses}
\begin{table*}[htbp]
\centering
\renewcommand{\arraystretch}{1.2}
\setlength{\tabcolsep}{5pt}
\begin{tabular}{|p{0.15\textwidth}|p{0.2\textwidth}|p{0.6\textwidth}|}
\hline
\textbf{Remote Associates Task} & \textbf{LLM Answer} & \textbf{LLM Guidance} \\
\hline
flake, mobile, cone (snow) & Snow (Snowflake, Snowmobile, Snowcone) &
Possible associated words for "flake" are snow, cereal, skin, coconut, and pastry.
Possible associated words for "mobile" are phone, sculpture, home, device, and car.
Possible associated words for "cone" are ice cream, volcano, traffic, pine, and speaker.
Now, try to identify any connections between these words and see if you can come up with a fourth word that is associated with all three. It might be helpful to jot down the words on paper to visually see any connections. Good luck! \\
\hline
dream, light, break (day) & Day (Daydream, Daybreak, Daylight) &
Possible associated words for "dream" are sleep, subconscious, nightmare, fantasy, and aspiration.
Possible associated words for "break" are fracture, pause, vacation, interruption, and opportunity.
Possible associated words for "light" are lamp, bright, illuminate, candle, and photon.
Take a moment to think about any connections between these words and try to come up with a fourth word that is associated with all three. Remember to jot down the words on paper if it helps you visualize any connections. Keep up the good work! \\
\hline
aid, rubber, wagon (band) & Band (Bandaid, Rubber Band, Bandwagon) &
Possible associated words for "aid" are help, support, assist, rescue, first.
Possible associated words for "wagon" are cart, horse, transport, carry, goods.
Possible associated words for "rubber" are elastic, tire, eraser, latex, ball.
Now, can you identify any connections between these words? Is there a fourth word that could connect them? Remember, it might be helpful to jot down the words on paper and see if any connections jump out at you. \\
\hline
\end{tabular}
\caption{Sample LLM responses for the exposure rounds in the convergent thinking experiment.}
\label{tab:rat-responses}
\end{table*}

\end{document}